\documentstyle[preprint,aps]{revtex}
\tightenlines
\newcommand{\be}{\begin{equation}}
\newcommand{\ee}{\end{equation}}
\newcommand{\ba}{\begin{eqnarray}}
\newcommand{\ea}{\end{eqnarray}}

\begin{document}
\title{On the chiral and deconfinement phase transitions in
parity-conserving $QED_3$ at finite temperature}
\author{I.J.R~Aitchison$^a$\,and\,
C.D.~Fosco$^{b}$\thanks{On leave from Centro At\'omico Bariloche, 
8400 Bariloche, Argentina}\,
\\
{\normalsize\it
$^a$University of Oxford, Department of Physics,
Theoretical Physics, 1 Keble Road,\\ Oxford OX1 3NP,
United Kingdom}\\
{\normalsize\it
$^b$The Abdus Salam ICTP, Strada Costiera 11, 
34100 Trieste, Italy}}
\date{\today}
\maketitle
\begin{abstract}
We present  some results about  the  interplay between the chiral  and
deconfinement phase transitions in parity-conserving $QED_3$ (with $N$
flavours of  massless $4$ component  fermions) at  finite temperature. 
Following Grignani et al.~\cite{sem1,sem2}, confinement is discussed 
in terms of an effective Sine-Gordon theory for the timelike component 
of the gauge field $A_0$. But whereas in ~\cite{sem1,sem2} the fermion 
mass $m$ is a Lagrangian parameter, we consider the $m=0$ case and ask 
whether an effective S-G theory can again be derived with $m$ replaced 
by the dynamically generated mass $\Sigma$ which appears below $T_{ch}$, 
the critical temperature for the chiral phase transition. The fermion 
and gauge sectors are strongly interdependent, but as a first approximation 
we decouple them by taking $\Sigma$ to be a constant, depending only on 
the constant part $\tilde{A}_0$ of the gauge field. We argue that the 
existence of a low-temperature confining phase may be associated with 
the generation of $\Sigma$; and that, analogously, the vanishing of 
$\Sigma$ for $T > T_{ch}$ drives the system to its deconfining phase. 
The effect of the gauge field dynamics on mass generation 
is also indicated.
\end{abstract}
\bigskip
\newpage
\section{Introduction}\label{intro}
Quantum   electrodynamics  in    (2+1)-dimensions  (QED$_3$)  is    an
interesting  theoretical laboratory in which to  explore, in a simpler
environment,  some of  the fundamental  features   of more complicated
four-dimensional gauge theories, such as  chiral symmetry breaking and
confinement; it  may also  be  of direct   physical relevance in  some
condensed matter systems  ~\cite{nicks1}. In this  paper,  we shall be
concerned  with $N$-flavour   QED$_3$ at  finite   temperature, in its
parity invariant (four-component) form.  Our aim  is two-fold: to cast
some light on  the nature of the phase  transition associated with the
dynamical   generation    of   fermion  mass    (the   ``chiral  phase
transition''); and to  argue  for a  dynamical  interplay between  the
chiral phase transition and the confinement-deconfinement one.

Chiral symmetry breaking  (``csb''), or equivalently dynamical fermion
mass  generation, in   QED$_3$ at $T   \not=  0$ has been  extensively
studied  using Schwinger-Dyson   equations,   in the large  $N$  limit
~\cite{nicks1,nicks2,ijra1,ijra2,ijra3,dom}. Though the approximations
are admittedly   not fully controlled,  all calculations   show that a
fermion    mass  is  dynamically  generated   for  $T<T_{ch}$, thereby
spontaneously      breaking  the    chiral      symmetry.   But    the
Coleman-Mermin-Wagner    theorem   ~\cite{cole,merwag}  forbids    the
occurrence  of  an order  parameter  transforming  non-trivially under
chiral transformations in  this two-dimensional system. One  presumes,
therefore,  that the transition  at  $T =  T_{ch}$  is of the BKT type
~\cite{ber,kos} . However, to our knowledge there has been no explicit
demonstration of this  --   for  example, by mapping    (or  otherwise
relating) the  theory to one with  a known BKT transition, which could
then be identified with the csb transition in QED$_3$.

On the other hand, in a quite independent development, Grignani et al.
~\cite{sem1,sem2}  have  shown  that, when  chiral  symmetry  is  {\it
explicitly}   broken  by a  fermion  mass  $m$  in the Lagrangian, the
relevant effective  theory (at least in  a certain region of parameter
space) has a BKT transition at a temperature  $T_c$, which the authors
interpret as a confinement-deconfinement phase transition.

It is natural to ask, first, whether the analysis of ~\cite{sem1,sem2}
can be extended  to the case in  which the fermion mass is dynamically
generated, rather than   appearing explicitly in the  Lagrangian;  and
secondly, whether the resulting  effective theory leads us any  nearer
to an   understanding  of the  chiral   phase transition,  and  of any
possible connection between the two  transitions. The purpose of  this
paper is to make a first attempt at answering these questions.

It   will  be  helpful to     recall   the argument  of   Grignani  et
al.~\cite{sem1,sem2}.  According to them, confinement in QED$_3$ at $T
\not= 0$ can be  probed by an  Abelian  analogue of the  Polyakov loop
operator, namely
\begin{equation}
P_{\tilde{e}}(\vec{x}) = e^{i \tilde{e} \int^{\beta}_0 A_0 (\tau, \vec{x}) 
d\tau},
\end{equation}

\noindent  where $\beta  =  1/T \; (k_B  =   1)$, and $\tilde{e}
\not=$ integer $\times  e_{f}$  where $e_{f}$   is the  charge of  the
dynamical fermions in the theory. The  Euclidean path integral for the
partition    function   is  invariant   under  gauge   transformations
$A^{'}_{\mu} = A_{\mu} + \partial_{\mu} \chi$, where

\begin{equation}
\chi(\beta, \vec{x}) = \chi (0, \vec{x}) + 2 \pi n/e_{f} 
\ \ \ , \ \  n \,\in \, Z.
\end{equation}

The group of all gauge transformations modulo those which are strictly
periodic is $Z$, the additive  group of integers; this $Z$-symmetry is
a global symmetry. Under a $Z$-transformation,

\begin{equation}\label{ptrans}
P_{\tilde{e}} (\vec{x}) \rightarrow P_{\tilde{e}} (\vec{x}) \, 
e^{2 \pi i n \tilde{e} /e_{f}}.
\end{equation}

If  $Z$-symmetry is unbroken, then  $<P_{\tilde{e}}(\vec{x})> = 0$ and
the   system  confines    $\tilde{e}$  charges;   if   $Z$-symmetry is
spontaneously broken  (as it  can be, with  an  order parameter, being
discrete), then $<P_{\tilde{e}}(\vec{x})> \not= 0$ and the $\tilde{e}$
charges are deconfined.

As noted in  ~\cite{sem1,sem2}, it  is possible to  choose  a gauge in
which $A_0$ is independent of $\tau$. In this gauge,

\begin{equation}
 P_{\tilde{e}}(\vec{x}) = e^{i \tilde{e} \beta A_0 (\vec{x})}
\end{equation}

\noindent and $Z$-transformations are

\begin{equation}
e_{f} \beta A_0 (\vec{x}) \rightarrow e_{f} \beta A_0 (\vec{x}) + 2 \pi n,
\end{equation}

\noindent under which $P_{\tilde{e}}$ transforms as in (\ref{ptrans}).
The effective field theory  for $P_{\tilde{e}} (\vec{x})$ is  then the
effective   action for  the  static field   $A_0  (\vec{x})$, which is
obtained by integrating out the other degrees of freedom from the path
integral.  The resulting effective action is  in general non-local and
non-polynomial, but it must still obey the $Z$-symmetry periodicity:

\begin{equation}\label{strans}
S_{eff}[e_{f} \beta A_0 (\vec{x})] = S_{eff}[e_{f} \beta A_0 (\vec{x}) + 2 \pi n].
\end{equation}

A local approximation can be obtained by expanding $S_{eff}$ in powers
of derivatives of $A_0 (\vec{x})$ divided by the (Langrangian) fermion
mass $m$. In  such an expansion, the  effective  potential is obtained
from the fermion determinant in the  presence of a constant background
field $A_0$ -- and it has to be periodic as in (\ref{strans}).  It is,
one   might  think, very  likely   that  such  a   potential should be
essentially  of  ``Sine-Gordon''  type, and  indeed  this is  what the
authors of ~\cite{sem1,sem2} were able to show, provided that

\begin{equation}\label{mcond}
m \gg T, e_{f}^2 /12 \pi \ \  
\end{equation}

\noindent (see also section II.A below).  In this regime, the critical
behaviour of the  effective theory for  $A_0 (\vec{x})$ is that of the
2-D Sine-Gordon model, which has a line of critical points along which
a BKT transition  occurs, separating confining and deconfining regions
of parameter space. The critical temperature is

\begin{equation}\label{tc}
T_c = \frac{e_{f}^2 / 8 \pi}{1 + e_{f}^2 / 12 \pi m + \cdots}
\end{equation} 

\noindent up  to  one-loop order. Thus  the  confinement-deconfinement
transition has been shown to be of BKT type, by demonstrating that the
relevant effective field theory has this critical behaviour.

Consider now how  dynamical (fermion) mass  generation (dmg) might fit
into this picture. Our basic idea is to see if  it might be consistent
for the  dynamically generated mass $\Sigma$ to  play  the role of the
Lagrangian  mass $m$  in    the analysis  of   ~\cite{sem1,sem2}.   To
investigate  dmg,  we  shall  use  Schwinger-Dyson  (SD) equations  as
previously,  but here    with one  important difference:   we  wish to
safeguard the crucial invariance  (\ref{strans}).  We  shall therefore
split $A_0$ into  its  slowly varying $(\tilde{A}_0)$  and fluctuating
$(a_0)$  parts, and - as a first approximation -   
solve  the  SD  equations for   the fermion  and
$a_0$-field propagators in  the presence of the  uniform $\tilde{A}_0$
which, at  this stage,  plays the role  of  an external parameter, and
which may be ``large''.  The fermion self-energy equation will produce
a   dynamically  generated mass  $\Sigma$ for  every   set of external
parameters (in an appropriate region of parameter  space) and for each
$\tilde{A}_0$.  The low momentum components of  $A_0$ are then treated
as in ~\cite{sem1,sem2} with $m$ replaced by $\Sigma$; 
 for the analysis to  go through as before
we shall need  to find that $\Sigma$  does not depend  too strongly on
$\tilde{A}_0$, at least in  some  parameter region. This picture  will
only be   consistent if (c.f.  (\ref{mcond})  ) the generated $\Sigma$
satisfies the condition

\begin{equation}\label{conds}
\Sigma \gg T, e_{f}^2 / 12 \pi
\end{equation}

\noindent  with,    of course,  $T<T_{ch}$.  In  this    case,  we can
tentatively    associate the generation   of    fermion mass  with the
occurrence of a confining phase.

Of  course, the fermion  and gauge dynamics  of $a_0$ are coupled: the
vacuum  polarisation function will be the  bubble diagram evaluated in
the $\tilde{A}_0$-background, and using the dynamically generated mass
$\Sigma$, which  also depends on  $\tilde{A}_0$. We shall see  that in
the low-temperature confined phase, where  $A_0$ is fluctuating  about
the trivial minimum of  the  Sine-Gordon potential, the properties  of
the gauge  field propagator  differ little  from  those in the  normal
$(\tilde{A}_0   =    0)$ case.  However,    for   $A_0$ configurations
fluctuating  near  a  position  of  unstable  equilibrium for  the S-G
potential -- which may be associated with the onset of the deconfining
phase transition -- we find that an instability develops in the vacuum
polarisation, which threatens to destroy the non-zero gap solution and
so restore  chiral symmetry. There is  an  indication here, therefore,
for an  intimate connection   between the  chiral and    the confining
transitions. Unfortunately, our method does   not allow us to  explore
the behaviour close to the phase transitions in any detail.

The paper is organised  as follows. In Section  II we define the model
and   discuss  the  formalism  used  to  study    confinement and  csb
simultaneously.  In Section III  we calculate the vacuum polarisation in  
the presence of $\tilde{A}_0$. In Section IV we write down the self-
consistent gap equation in the presence of $\tilde{A}_0$, and present 
numerical results together with our conclusions. An Appendix discusses 
the error due to  neglecting the dynamics of the  space  components of
the gauge field.

\section{Towards a unified treatment of the chiral and \\
deconfinement phase transitions}\label{uni}

In this  work we shall  be  concerned with  a  theory whose generating
functional       of     complete     Green's       functions    ${\cal
Z}[j_\mu;{\bar\eta},\eta]$,   in   the imaginary     time  (Matsubara)
formalism, is defined by
\begin{equation}\label{defz}
  {\cal Z}[j_\mu;{\bar\eta},\eta] \;=\; \int \, {\cal D}A_\mu {\cal D}
  {\bar\psi}{\cal D}{\psi}\, \exp \left\{- S + \int d^3x [j_\mu(x)
  A_\mu(x) + {\bar\eta}(x) \psi(x)
  + {\bar\psi}(x) \eta(x) ] \right\}
\end{equation}
where $\int d^3x \cdots \equiv \int_0^\beta d\tau \int d^2x \cdots$,
$\beta =\frac{1}{T}$, and $S$ denotes the Euclidean action
 \begin{equation}\label{defs}
  S\;=\;\int d^3x \,{\cal L} \;\;\;\;\;
  {\cal L} \;=\; {\cal L}_G + {\cal L}_F \;.
  \end{equation}
The gauge field Euclidean Lagrangian, ${\cal L}_G$, is assumed to be
 of the standard Maxwell type,
\begin{equation}\label{deflg}
  {\cal L}_G \;=\; \frac{1}{4} \, F_{\mu\nu}F_{\mu\nu}
  \;\;\;\;\;\;\, F_{\mu\nu}\;=\;\partial_\mu A_\nu - \partial_\nu A_\mu \;,
\end{equation}
where we  have omitted,  for the time  being, the  gauge fixing.   The
fermionic    field $\psi$ denotes   $N$ flavours   $\psi_a$, each  one
minimally coupled to    the gauge  field $A_\mu$.  The   corresponding
Euclidean Lagrangian, ${\cal L}_F$, is then
\begin{equation}
 {\cal L}_F \;=\; \sum_{a=1}^N {\bar\psi}_a(\tau,x)  \left(
\not \!\partial + i e_N \not\!\! A + m \right)
 \psi_a (\tau,x)
\end{equation}
where the coupling constant $e_N  = \frac{e}{\sqrt{N}}$ (replacing the
$e_{f}$ of Section I) has the proper $N$-dependence to ensure that, in
the   large-$N$ limit   keeping $e$   fixed,  all  Feynman diagrams  are
finite. The Lagrangian mass $m$ may or may not be present.
We use  the notation $\tau$ and $\vec  x$ to refer to the time
and space coordinates, respectively. The $\gamma$-matrices are in a $4
\times 4$  (reducible)  representation of the  Dirac  algebra in $2+1$
dimensions:
\begin{equation}
\gamma_\mu \;=\; \left( \begin{array}{cc}
                         \sigma_\mu & 0 \\
                          0 &- \sigma_\mu
                         \end{array} \right) \;,
\end{equation}
where $\mu \,=\, 0,1,2$, and
\begin{equation}
\sigma_0 \;\equiv\; \left( \begin{array}{cc}
                         1 & 0 \\
                          0 &- 1
                         \end{array} \right) \;\;
\sigma_1 \;\equiv\; \left( \begin{array}{cc}
                          0 & 1 \\
                          1 & 0
                         \end{array} \right) \;\;
\sigma_2 \;\equiv\; \left( \begin{array}{cc}
                          0 & -i\\
                          i &  0
                         \end{array} \right) \;,
\end{equation}
are the usual  Pauli  matrices.  This representation allows   for  the
introduction of a parity-conserving  mass term for  the four-component
fermions. This corresponds, in   the two-component formalism, to  mass
terms with   opposite signs  for   each fermion  field within a  given
flavour. We shall here use  a description that allows  us to study the
possibility   of both    chiral  and deconfinement  phase  transitions
simultaneously. Note that  the latter   transition, as presented  in 
~\cite{sem1,sem2}, is considered in terms of an {\em effective
action\/}   for  $A_0$, the time component    of the gauge field. This
effective action  is derived in  the static-$A_0$  gauge, performing a
derivative  expansion.   The usual   way  to study   the  chiral phase
transition   is,   on  the   other  hand,   to  work    with truncated
Schwinger-Dyson equations (SDE's), an approximation which is justified
in the  large-$N$ limit.  It is  possible, as we shall  see, to find a
non-empty region in parameter space  where the simplifying assumptions
made for the study of each phase transition are indeed compatible.

 We shall briefly review here  the derivation of the effective  action
for  the field $A_0$, for  the case of  massive fermions, following 
~\cite{sem1,sem2}. This effective action  is the interesting  object
to study when dealing with the deconfinement phase transition. We will
afterwards   extend   this  procedure,   in  order    to  include  the
non-perturbative  dynamics   responsible   for   the    chiral   phase
transition.  The  fermion   mass,  rather   than being  a   Lagrangian
parameter, will then be dynamically generated, and determined by a gap
equation. Following  references~\cite{sem1,sem2} (with trivial changes
due to the fact that we now have $N$ fermionic flavours) the effective
action  $S_{eff}[A_0]$ for  a  system where   the  fermions do have  a
Lagrangian mass $m$, is obtained by first using a gauge transformation
such that $A_0$ is $\tau$-independent~\footnote{$A_0$ itself cannot be
set equal to zero, because the time  coordinate is compact.}, and then
integrating out all the remaining fields, i.e.,
\begin{equation}\label{sa}
 e^{-S_{eff}[A_0]}\;=\;
 e^{-\frac{1}{2} \int d^3x (\partial_jA_0)^2}
 \int {\cal D}{\vec A}\,
 \exp\left\{ -\int d^3x [
                          \frac{1}{2} (\partial_0 A_j)^2
                         +\frac{1}{4} F_{jk}^2 ]\right\}
 \times\, e^{-\Gamma_N[A_0,A_j]}
\end{equation}
where the functional $\Gamma_N[A_0,A_j]$ is the result of integrating out
the fermionic fields,
$$
\Gamma_N [A_0,A_j]\;=\;\int\,
             \Pi_{a=1}^N {\cal D}{\bar\psi}_a {\cal D}\psi_a\,
      e^{- \int d^3x
        \sum_{a=1}^N
        {\bar\psi}_a
        [\not \partial + i e_N \gamma_0 A_0 + i e_N \gamma_j A_j
        + m ] \psi_a}
$$
\begin{equation}\label{seff}
=\; \left\{ \det[ \not \!\partial + i e_N \gamma_0 A_0(x) + i e_N
 \gamma_j A_j (\tau,x) + m ] \right\}^N \;.
\end{equation}
Of course,    we have $\Gamma_N  [A_0,A_j]\,=\,N  \,\Gamma [A_0,A_j]$,
where $\Gamma [A_0,A_j]$ denotes the contribution corresponding to the
integration    of   just    one     fermionic   flavour.     Following
\cite{sem1,sem2}, $\Gamma[A_0,A_j]$   may   be evaluated by   using  a
derivative expansion for the external gauge field, taking into account
the fact that    $A_0$ may have `large'  values,   i.e.,  its constant
component cannot be assumed  to be small,  since that would spuriously
break  the invariance under shifts  in $A_0$.  To use an approximation
which preserves this  periodicity  is crucial
 for the study of the  deconfinement phase transition.  Thus
we let
\begin{equation}\label{aexp}
  A_\mu (x) \;=\; {\tilde A}_\mu (x) + a_\mu(x)
\end{equation}
where  ${\tilde A}_0$  is  the large, almost  constant piece, ${\tilde
A}_j =  0$,  and $a_\mu(x)$  denotes the  fluctuating   part of $A_\mu
(x)$. As ${\tilde  A}_0$ cannot be  assumed to be small, the expansion
of the determinant must include that object exactly:
$$
\det[\not \!\partial + i e_N \gamma_0 A_0(\vec x) + i e_N \gamma_j A_j
(\tau,\vec x) + m ]\;=\;\det[\not  \!\partial + i e_N \gamma_0 {\tilde
A}_0 + m ]
$$
\begin{equation}\label{detex1}
\times \,\det\left[1 + i e_N(\not \!\partial +  i e_N \gamma_0 {\tilde
A}_0 + m)^{-1} \not \! a (\tau,\vec x)\right] \;,
\end{equation}
where, for  the  last  factor  in  (\ref{detex1}),  we can  attempt  a
small-$a_\mu$   expansion.  In   a first  approximation,    and to  be
consistent with   the  approximation  usually invoked   in   the SDE's
approach to the study of the chiral phase  transition, we shall ignore
the dynamics  of $a_j$, the  space components  of  the gauge  field. A
discussion  about the error due  to this approximation is postponed to
the Appendix.   Then,  we only  need   the fermionic  determinant as   a
function of the time component of the gauge field. For one flavour, we
have
$$
e^{-\Gamma [A_0]}\;=\; \det [\not\!\partial + i e_N \gamma_0 A_0 (\vec
x) + m ]
$$
\begin{equation}\label{detex2}
\;=\;\det [\not\!\partial   + i e_N \gamma_0   {\tilde A}_0  + m  ] \;
\det\left[1 +  i e_N(\not  \!\partial  + i  e_N \gamma_0{\tilde A}_0 +
m)^{-1} \gamma_0 a_0 (\vec x)\right] \;.
\end{equation}
 The $a_0$-independent   factor   in  (\ref{detex1}),  involving    no
derivatives   of the  gauge field,  plays  the role   of an  effective
potential, and is given by
\begin{equation}
\det [\not\!\partial + i  e_N \gamma_0 {\tilde  A}_0 + m ]  \;\equiv\;
e^{\int d^2x V (m, \frac{e_N {\tilde A}_0}{T})}
\end{equation}
with
\begin{equation}
V (m,   \frac{e_N   {\tilde A}_0}{T})\;=\;   \frac{1}{Vol.}    \log \det[   (-i
\partial_0 - e_N {\tilde A}_0)^2 - \nabla^2 + m^2 ]\;.
\end{equation}
Regarding the  second  factor,  when  expanded for  low  momentum,  it
produces   a wave function  renormalization  of  the  kinetic term for
$A_0$.  Putting together  these  two pieces, and  including the factor
$N$, we see that the effective action for $A_0$ becomes
\begin{equation}
S_{\rm    eff}[A_0]=\int d^2  x \left(     Z(m,e_N {\tilde A}_0/{{T}})
\frac{1}{2T}\vec\nabla  A_0\cdot\vec\nabla   A_0 \,-\,   N\,   V(m,e_N
A_0/{T}) \right)\,,
\end{equation}
where
\begin{equation}
N    \,  \Pi_{00}(0,\vec k^2)\;   =    \;  N  \,  \Pi_{00}(0,0)+  \vec
k^2(Z-1)+\ldots
\end{equation}
where $\Pi_{00}(n,\vec k^2)$ denotes the  $00$ component of the vacuum
polarization  tensor in   the presence  of the    constant gauge field
${\tilde A}_0$.

To   relate  this  form of  the  effective   action  to  a Sine-Gordon
description,  one performs   a   harmonic expansion of  the  effective
potential:
\begin{equation}\label{v1}
V(m,e_N    A_0/{T})=-\frac{T^2}{\pi}\sum_{n=1}^{\infty}(-1)^n     {\rm
~e}^{-n    m/T}       \left(1+\frac{n   m}{T}\right)\frac{\cos(n   e_N
A_0/{T})}{n^3}\;.
\end{equation}

This effective potential may  be further simplified in some particular
regimes.  For example, when $m>>T$, $T/m$ and  $e_N^2/(12 \pi m)$ are small and
$e_N^2/T$ is arbitrary, so that $Z \approx 1$ and the higher harmonics
in the effective potential are  exponentially small perturbations; one
may then keep only the leading term,
\begin{equation}\label{v2}
V(m,e_N  A_0/T)\,\sim \,\frac{T    m}{\pi}{\rm      ~e}^{-m/T}\cos(e_N
A_0/{T})\ \ ,
\end{equation}
which  is a sine-Gordon potential.  Under  these conditions, one has an
effective description which corresponds to the Sine-Gordon theory for the 
scalar field $A_0$ : 
\begin{equation}\label{ssg}
S_{\rm  eff}^{(m>>T)}[A_0]=\int d\vec x\left\{ \frac{1}{2T} \vec\nabla
A_0\cdot\vec\nabla  A_0-  N  \frac{T  m}{\pi}{\rm ~e}^{-m/T} \cos( e_N
A_0/T)\right\},
\label{aux1}
\end{equation}
which has a BKT~\cite{ber,kos} phase transition.
The  ``$\beta$''   parameter of  the   Sine-Gordon  model is  now  $N$
dependent,  and given by  $\beta  = \frac{e}{\sqrt{N T}}$.  Under  the
assumption  (\ref{mcond})   (with $e_{f}$  replaced   by $e_{N}$), the
critical  temperature is given  by (\ref{tc}) with $e_{f}$ replaced by
$e_{N}$.

It is important to remark that, in this analysis, the Lagrangian mass $m$
has been regarded  as a free parameter.  We shall now argue that, when
$m$ is zero, one can - under certain conditions - arrive at an effective
theory of  the  same form  as  (\ref{aux1}), but with  $m$ replaced by
$\Sigma$, the dynamically induced gap.

The effective action (\ref{ssg})  
has  been  obtained  by  performing  a derivative
expansion.  In  consequence,  when the   $A_0$ field  is   regarded as
dynamical,  the effects due  to  its large  momentum modes are  either
neglected, or improperly taken into account.  This  is, of course, not
an issue when one is dealing with  the deconfinement phase transition,
but  it will have an  important influence for   the particular case of
{\em massless\/} fermions.  In this  case the  mass parameter used  to
perform the low-momentum expansion is proportional to the temperature,
and one is then missing modes of  the gauge field  which may indeed be
responsible  for    the  generation  of  a   dynamical   mass  for the
fermions. In  other words,  the  procedure  of using the  low-momentum
effective action  for $A_0$ for the case  of  massless fermions would 
not be 
accurate, since  one still has   to include virtual  $a_0$ corrections
which contribute to the  generation of a  mass for the fermions.  If a
non-vanishing  dynamical mass    for  the fermion    is generated, the
momentum expansion procedure can be made consistent, but one should be
careful to   include  in that   dynamical mass  the  dependence on the
temperature and on the  parameters of the  model. 
 
 We shall now   give  a formal derivation  to  justify  this
procedure.  The definition of the effective action for the large field
$A_0$ should now take into account  the existence of the high momentum
modes  for  $A_\mu$. The natural way  to  proceed is  to decompose the
gauge field  into its large and small  components, as in (\ref{aexp}),
where only up to  two derivatives of  ${\tilde A}$ will be kept, while
$a_\mu$ is assumed to take care  of the higher derivatives.  We shall,
to simplify the notation,  get rid of  the tilde, and $A$  will denote
just   the large  component of  the  gauge  field. Then,  $S_{eff}$ is
defined by 

\begin{equation}
\exp\{-S_{eff}[A_0]\}= \int {\cal D}a_\mu  \Pi_a {\cal D}{\bar \psi}_a
{\cal D}\psi_a \exp\{-S[A_0 + a_0, a_j,{\bar\psi},\psi]\}
\end{equation}
where $S$ is the Euclidean action, as defined in (\ref{defs}). More
explicitly,
\begin{equation}\label{saex}
S[A_0 + a_0, a_j,{\bar\psi},\psi]= \int d^3 x \left[ 
\frac{1}{4} F_{\mu\nu}(A+a)F_{\mu\nu}(A+a)
+ \sum_b {\bar\psi}_b (
\not \!\partial + i e_N (\not\!\! A + \not \! a) )
 \psi_b \right]\;.
\end{equation}
It is crucial to realise that $A$ and $a$ are assumed to have
complementary supports in momentum space. Thus the mixed quadratic
term for the gauge field in (\ref{saex}) vanishes
\begin{equation}
\int d^3 x \, \frac{1}{2} F_{\mu\nu}(A) F_{\mu\nu}(a) = 0 \;,
\end{equation}
as can be easily verified by Fourier transformation.

Then we see that
\begin{equation}
\exp\{-S_{eff}[A_0]\}\,=\,\exp \{-\frac{1}{2} \int d^3x (\partial_jA_0)^2
- W[A_0]\}
\end{equation}
where
\begin{equation}\label{efd}
e^{-W[A_0]} = \int {\cal D}a_\mu {\cal D}{\bar \psi} 
{\cal D}\psi \exp\{- \int d^3 x [
\frac{1}{4} F_{\mu\nu}(a)F_{\mu\nu}(a)
+ \sum_{b=1}^N {\bar\psi}_b \left(
\not \!\partial + i e_N (\not\!\! A + \not\! a) \right)
 \psi_b ]\}\;.
\end{equation}
Note that $W[A_0]$ has a simple interpretation: it is the free energy
corresponding to parity conserving $QED_3$ in the presence of an
$A_0$ background. The leading term in a large $N$ expansion is 
\begin{equation}\label{sf}
W[A_0] \,=\, - {\rm Tr} [ \ln S_F^{-1} ]
\end{equation}
where $S_F$  denotes   the full  fermion   propagator, for  the theory
defined by  (\ref{efd}). This propagator is the  result of solving the
S-D  equations in the $A_0$   background.

In principle, since $A_0$ depends on $x$ so will $\Sigma$, but we shall 
make the assumption that the induced mass is not sensitive to the 
gradient of $A_0$, but only to the constant part $\tilde{A}_0$. This 
assumption will only be reliable in the confined phase, where the 
topological excitations of the S-G theory are absent, and $A_0$ 
has only small fluctuations around one of the stable minima of 
the effective potential. With $\Sigma$ assumed constant, $S_F^{-1}$ 
will have the form 
\begin{equation}\label{vac}
S_F^{-1} = \not \!\partial + i e_N \gamma_0 A_0 + \Sigma \;;
\end{equation}
evaluation of the trace in  (\ref{sf}) may then proceed as in the 
$m \neq 0$ case, via Equations (\ref{detex2}) - (\ref{ssg}) above. But an  
important question to answer will be whether, for at least some part 
of the solution-space for $\Sigma$, conditions (\ref{conds}) do in 
fact hold, allowing the passage from (\ref{v1}) to (\ref{v2}). 

The low-momentum effective action for the gauge field is then 
constructed with the 1PI functions derived (see Sections III and IV) 
from the self-consistent SD equations, with $\tilde{A}_0$ appearing 
as an external field. As the low-momentum component of the gauge 
field has not yet been integrated, one has to treat its dynamics 
separately; this is done with the S-G description, as in~\cite{sem1,sem2}, 
which does indeed allow for a BKT transition.  
The phase transition for the deconfinement transition should then be 
studied taking into account the fact that $\Sigma$ is now not an 
independent parameter (as was $m$), but rather a function of $T$, 
$e$, $N$ and $\tilde{A}_0$. In this connection, we shall have to 
check that the dependence of $\Sigma$ on $\tilde{A}_0$ is, in fact, 
not so rapid as to change the effective potential in (\ref{ssg}) 
significantly away from the S-G form.

\section{The vacuum polarisation function $\Pi_{00}$ in the presence 
of $\tilde{A}_0$}\label{pol}
\vspace{1cm}

We derive here the vacuum polarisation function in the one-loop (large-$N$)  
approximation, keeping the full dependence on the `large' (uniform) 
piece of the time-like component of the gauge field, $\tilde{A}_0$.

In principle, the coupled S-D equations for the 1PI functions 
should be solved self-consistently, having made the usual truncations. 
However, experience shows ~\cite{pis} that, in dynamical mass 
calculations, it makes very little difference whether a fermion 
mass is included in $\Pi_{00}$ or not. We shall present the 
calculation for massless fermions, indicating at the end the 
changes needed to deal with massive ones.

For  one  single
four-component fermionic flavour  (for  $N$ flavours we only   have to
multiply the result  by $N$), the object  we consider is the quadratic
form $\Gamma^{(2)}[A_0]$, defined by
\begin{equation}\label{defpi}
\Gamma^{(2)}[A_0] \;=\; \frac{1}{2\beta}\, \int \frac{d^2k}{(2\pi)^2}\,
{\tilde a}_0(0,-k) \Pi_{00}(0,k) {\tilde a}_0(0,k)
\end{equation}
where ${\tilde a}_\mu(n,k)$ denotes the Fourier transformed of $a_\mu$,
namely
\begin{equation}
a_\mu(\tau,x)\;=\;\frac{1}{\beta} \sum_{n=-\infty}^{+\infty} e^{i
(\alpha_n \tau + k \cdot x)} {\tilde a}_\mu(n,k)
\end{equation}
with $\alpha_n \,=\, \frac{2\pi n}{\beta}$, the bosonic Matsubara
frequency. From the definition of the fermionic determinant, we
see that the vacuum polarization tensor is given by
\begin{equation}\label{defpi1}
  \Pi_{\mu\nu}(n,k) \;=\; - \frac{e_N^2}{\beta} \sum_{l=-\infty}^{l=+\infty}
  \int \frac{d^2p}{(2\pi)^2} {\rm tr}
   \left[
  \frac{1}{i (\not\!p+\not\!k)+ i \gamma_0 {\tilde \omega}_{l+n}}
  \gamma_\mu 
   \frac{1}{i \not\!p+ i \gamma_0 {\tilde \omega}_l} \gamma_\nu
   \right]
\end{equation}
where we adopted the notation ${\tilde \omega}_l  \,=\, \omega_l + e_N
{\tilde A}_0$,  where  ${\omega}_l=(2 l   +1)\frac{\pi}{\beta}$ is the
fermionic   Matsubara frequency.  Note  that the   Dirac slash is  now
defined for  the two spacelike  indices in the `dimensionally reduced'
theory. The `${\rm tr}$' denotes Dirac  trace, over the four component
spinor space.   After evaluating this  Dirac trace, and performing the
angular    integration   for the    spatial    momentum,   $F(k)\equiv
\Pi_{00}(0,k)$,  the zero frequency  part of $\Pi_{00}$ may be written
as follows
\begin{equation}\label{fex}
F(k)\;=\; e_N^2 \int_0^1 d\alpha \, \int_0^\infty \frac{dp^2}{\pi} \,
\frac{1}{\beta} \, \sum_{n=-\infty}^{+\infty}
 \frac{p^2-\alpha (1-\alpha) k^2 -
 {\tilde \omega}^2_n}{[p^2+\alpha (1-\alpha)k^2 +
 {\tilde \omega}^2_n]^2} \;,
\end{equation}
where $\alpha$ is a Feynman parameter.  To  calculate $F(k)$, we found
 it convenient   to split that  function  into $F_{T=0}(k)$ (its $T=0$
 limit) and  ${\bar  F}(k)= F(k)  -F_{T=0}(k)$. By  performing the sum
 over the discrete frequencies and the integration over the modulus of
 the  spatial momentum, the  subtracted  function ${\bar F}(k)$ may be
 expressed as an integral over $\alpha$:
\begin{equation}\label{fbar}
 {\bar F}(k) \;=\;  \frac{e_N^2}{\pi\beta}\, \int_0^1 d\alpha
 \ln [1+2 \cos (e_N \beta {\tilde A}_0)
 e^{-\beta \sqrt{\alpha (1-\alpha)} k}
 +e^{-2 \beta \sqrt{\alpha (1-\alpha)} k}] \;.
\end{equation}

\noindent The zero-temperature value of $F$ is well-known 
~\cite{pis}, and given by $\alpha_{N} k/8$.

It  does   not  seem possible  to    evaluate  (\ref{fbar})  in closed
form.  However, since  $\Pi_{00}$   appears  inside an  integral   in 
the equation for the dynamically generated fermion mass (Eq.(\ref{sdeq})
below),  it would  make the   solution of (\ref{sdeq})   much
easier   if an  accurate  analytic approximation   were  available for
(\ref{fbar}).  A similar problem was encountered in ~\cite{ijra1}, and
there a convenient approximation was  found which preserved  correctly
the  limits $\beta k \rightarrow  0$ and  $\beta k \rightarrow \infty$
(see Eq.(9) of ~\cite{ijra1}). Following  the same procedure here,  we
approximate $\Pi_{00}$ by

\begin{equation}\label{fapp}
\Pi_{00}  (0,k) \approx \frac{e^2}{8 \beta}  \left[  \beta k + \frac{8
\ln  [2 + 2  \cos  (e_{N} \beta \tilde{A}_0  )]}{\pi}  \exp \left\{ {-
\frac{\pi \beta k}{8  \ln [2 +  2  \cos (e_{N} \beta \tilde{A}_0  )]}}
\right\} \right]
\end{equation}
which incorporates

\begin{equation}
\lim_{\beta k \rightarrow \infty} \Pi_{00}(0,k) = \frac{e^{2} k}{8}
\end{equation}
and

\begin{equation}\label{mth}
\lim_{\beta  k \rightarrow 0} \Pi_{00}(0,k)  = \frac{2 e^2}{\pi \beta}
\ln \left\{ 2 \left\vert \cos \left( \frac{e_{N} \beta \tilde{A}_0}{2}
\right) \right\vert \right\},
\end{equation}

\noindent and where we  have now written  the result explicitly for $N
\neq 1$ flavours (recall   that $e_{N}=e/\sqrt{N}$, and $e^2$  is held
fixed as $N$ varies).  Expression (\ref{mth}) is the ``thermal mass'',
in the presence of  $\tilde{A}_0$.  It is clear  that when $e_N  \beta
\tilde{A}_0 >   \frac{2  \pi}{3}$ the    thermal mass  goes  negative.
We shall discuss this further in Section IV. For the moment, note that 
this value is in any case approaching near to 
 the ``unstable vacuum'' of  the S-G potential at
$e_N     \beta \tilde{A}_0  =   \pi$, which lies beyond the region 
of validity of our approach, based as it is on perturbing around one 
of the stable minima of the S-G potential. 
   For   $e_N   \beta \tilde{A}_0
\stackrel{<}{\sim} 1.5$, we  find that  the approximation (\ref{fapp})
is always within 10\% of the exact formula (\ref{fbar}).

It is not hard to show that, when $m\neq 0$, the zero frequency
part values of $\Pi_{00}$ are given by a function $F(k)$ defined
by
\begin{equation}
  F(k)\;=\; e_N^2 \int_0^1 d\alpha \, \int_0^\infty \frac{dp^2}{\pi} \,
\frac{1}{\beta} \, \sum_{n=-\infty}^{+\infty}
 \frac{p^2-\alpha (1-\alpha) k^2 -
 {\tilde \omega}^2_n + m^2}{[p^2+\alpha (1-\alpha)k^2 +
 {\tilde \omega}^2_n + m^2]^2} \;,
\end{equation}

An analogous procedure to the one followed in the massless fermion
case allows us to write the subtracted function ${\bar F}$ as
\begin{equation}\label{fbar1}
 {\bar F}(k)  = \frac{e_N^2}{\pi\beta}\, (1-m \frac{\partial}{\partial
 m}) \int_0^1   d\alpha  \ln [ 1+2    \cos (e_N  \beta   {\tilde A}_0)
 e^{-\beta     \sqrt{\alpha       (1-\alpha)k^2+m^2}}   +e^{-2   \beta
 \sqrt{\alpha(1-\alpha)k^2+m^2}}] \,.
\end{equation}
As for the massless case, an  approximate expression for ${\bar F}(k)$
may also be obtained,  but in our  calculations  we have not used  the
massive form of $\Pi_{00}$. In a fully self-consistent treatment, $m$ 
would be replaced by $\Sigma$ in (\ref{fbar1}).

\section{Dynamical mass generation in the presence of 
${\tilde A}_0$}\label{res}
\vspace{1cm}

In this first exploration of the interplay between csb and confinement
in QED$_3$ at $T \not= 0$, we shall adopt the simplest approach to the
S-D  equation for $\Sigma$ (the  ``gap equation'')  which was taken in
the  early paper  by Dorey and  Mavromatos ~\cite{nicks2}  . We  shall
assume  that   $\Sigma$  is momentum-independent,   that  fermion wave
function renormalization may  be   neglected,   and that   only    the
instantaneous-exchange  part     of  the  kernel  is     retained (see
~\cite{ijra3,dom} for evidence that   relaxing these assumptions  will
not  change the  qualitative    conclusions very much). The new feature 
of our calculation of $\Sigma$, as for $\Pi_{00}$, is the inclusion 
of the $\tilde{A}_0$ background.

The fermion propagator has the  form 
$(i \not \! p  + i \gamma_0 \tilde{\omega}_n
)^{-1}$ where $\tilde{\omega}_n  =  \omega_n + e_N   \tilde{A}_0$ and 
$\omega_n$ is the fermionic Matsubara frequency as before; the 
 gap equation is then 

\begin{equation}\label{sdeq}
1  =  \frac{e^2}{4 \pi N}   \int \frac{kdk}{[k^2  + \Pi_{00}(k, \beta,
\tilde{A}_0)]   \sqrt{k^2    +  \Sigma^2}} \left[   \frac{\sinh  \beta
\sqrt{k^2 +  \Sigma^2}}{\cosh  \beta \sqrt{k^2  + \Sigma^2}  + \cos (e
\beta \tilde{A}_0 / \sqrt{N})} \right ]
\end{equation}

\noindent  where $\Pi_{00}$ is the $\mu  = \nu =   0$ component of the
vacuum  polarization tensor $\Pi_{\mu \nu}$  as calculated in the 
preceding Section, Eq. (\ref{fapp}), 
and $e^2$ is fixed as $N$ varies. Eq.(\ref{sdeq}) is to be compared
with Eq.(20) of ~\cite{nicks2}, to  which it reduces when $\tilde{A}_0
\rightarrow 0$ (note that the latter equation is  in error by a factor
of $\frac{1}{2}$  on the  right  hand  side: ``$2N\pi$''  should  read
``$4N\pi$'', so that the $N=2$ results of ~\cite{nicks2} correspond to 
our $N=1$ results,etc.; 
  note also  that our  $\Pi_{00}$  is equal to $k^2 \times
\Pi_{00}$   of ~\cite{nicks2},  but   is    the same  as   $\Pi_0$  of
~\cite{ijra1}). The upper limit of integration in (\ref{sdeq}) is taken 
to be unity, as in ~\cite{nicks2}.

 It is worth noting that the momentum  integral in the S-D equation for
$\Sigma$ should really have an IR cutoff $\epsilon$,  given by a small
fraction of $\Sigma$, since the gauge field  corresponds to the
{\em fluctuating\/} part  of $A_0$. This  should not make any difference
to  the solution of  the equation, for all the  cases where there is a
non-vanishing $\Sigma$, since there the IR  behaviour of the integrand
is smooth.  We have verified numerically that indeed the solution to 
(\ref{sdeq}) is insensitive to the introduction of a small infrared cutoff.

On  the other  hand,  when  the only  stable  solution is
$\Sigma=0$, the   IR cutoff  would  make  a  difference, but  then the
separation into low   and high momentum  modes collapses.   Again,this
would correspond  to the physics  near to   the transition, where  our
procedure is not applicable.

We turn now to a discussion of the numerical solutions of (\ref{sdeq}), 
and their implications. 
Figure 1 shows $s = \frac{\Sigma}{e^2}$ as a function of $t = 
\frac{T}{e^2}$ for various values of $b = \frac{{\tilde A}_0}{e}$, 
for the case $N = 1$. For $b = 0$, the curve reproduces the known results 
of ~\cite{nicks2}. For $b \neq 0$, a mass continues to be 
generated; indeed, at a given temperature the mass increases with 
$b$ (see Figure 2 where $s$ is shown as a function of $b$ for 
fixed $t$). However, for $b \neq 0$ we cannot probe temperatures below
$\frac{3 b}{2 \pi}$, since below this value the thermal mass goes 
negative, as remarked after (\ref{mth}). In fact, from 
(\ref{fapp}) it is clear that when $t = \frac{3 b}{2 \pi}$ (i.e. 
$e \beta {\tilde A}_0 = \frac{2 \pi}{3}$), $\Pi_{00}$ reduces to 
$\frac{e^2 k}{8}$, the zero temperature value. This is the reason why the 
curves all tend to  $s(t=0)$ as $t$ 
tends to  $t=\frac{3 b}{2 \pi}$. 

In reality, the calculation should probably only be accepted for $t \geq 
\frac{b}{\pi}$ (i.e. $e\beta {\tilde A}_0 \leq  \pi$), 
since we must remember that, at $e\beta {\tilde A}_0 = \pi$, the Sine-Gordon 
potential in (\ref{aux1}) has a point of unstable equilibrium, while our mass
calculation is understood as being performed around a stable minimum of the
potential.
 Inspection of Figure 1 shows that the condition $t \geq 
\frac{b}{\pi}$ is met for the curves corresponding to 
$b = 0.01$ and $0.02$, but only for part of the 
$b = 0.03$ curve, and for less of the $b = 0.04$ curve.

 Figure 2 shows $s$ as a function of $b$ for various values of $t$, 
also for the case $N=1$. As before, 
 the curves tend to  the zero-temperature value as $b$ tends to  
$ \frac{2 \pi}{3} t $.  For the values of $t$ and 
$b$ shown, the more conservative limit $b \leq \pi t$ is 
always satisfied. 
The critical temperature for $b = 0$ is near  
$t_{ch}=0.0137$, which is why the curve for this value of $t$ ``divides'' 
the curves in Figure 2. 

We must now consider whether these results support our approach {\it a 
posteriori}. First and most importantly, we need to check whether 
conditions (\ref{conds}) hold. Inspection of Figure 1 shows that indeed 
the condition $\Sigma >> T$ (i.e. $s >> t$) 
is met for small $T$, say $T \leq 0.01$.
The condition $s >> \frac{1}{12 \pi N}$ is less satisfactorily met,
but it is not flagrantly violated. Secondly, we see from Figure 2 that, 
while there is certainly a dependence of $\Sigma$ on $\tilde{A}_0$, it 
does not seem to be sufficiently rapid to call into question the 
effective S-G description of the deconfinement transition.

So far we have discussed the case $N=1$, and we must now remember that 
some approximations we have made depend on the large-$N$ limit. However, it 
is well known (see for example ~\cite{ijra1} and references therein) 
that in the model considered here there is a critical value of $N$,
$N_{c}(T)$, above which chiral symmetry is restored. $N_{c}$ is 
typically in the region 1-3. We shall therefore consider the case 
$N=2$ for definiteness. While $N=2$ is hardly a large value, there is 
evidence to suggest that the inclusion of terms which are  higher order in 
$1/N$ will not alter the qualitative conclusions ~\cite{nash}.

 In Figure 3
we present, for  $N = 2$ , the curves of $s$ as a function of $t$, 
for two values of $b$. We see that both the critical temperature and the
value of $s$ at $t=0$ are reduced, but still there is an important fraction of
parameter space where $s >> t$. As for the condition 
$s >> \frac{1}{12 \pi N}$, it is satisfied to the same extent as for the $N=1$ 
case, since going from $N=1$ to $N=2$ reduces both sides of that inequality by
(roughly) a factor  $2$.

We may conclude that we have demonstrated the feasibility of our 
approach, in which the dynamically generated $\Sigma$ replaces (at
least in a limited region of parameter space) the Lagrangian mass $m$ 
of ~\cite{sem1,sem2}. We may therefore argue that in the $m=0$ case, 
dynamical generation of $\Sigma$ is associated with a low-temperature 
confining phase, via the emergence of an effective S-G theory. There 
is also an indication of a (reciprocal) effect of the gauge field 
dynamics on mass generation. As noted after Eq. (\ref{mth}), when 
$e_{N} \beta \tilde{A}_0 > (2 \pi /3)$ the thermal mass goes negative, 
which would imply a singularity in (\ref{sdeq}) due to the 
ensuing pole in the gauge field propagator in the Euclidean region. 
We may interpret this as indicating that, as the gauge field fluctuates 
towards the value $e_{N} \beta \tilde{A}_0 = \pi$, which is 
an unstable stationary point of the S-G potential, the stability of 
the gap equation is lost, and mass generation ceases. Interestingly, 
apparently similar special temperatures occur in a study of the Gross-Neveu 
model with an imaginary chemical potential~\cite{petkou}.
Clearly it would be desirable to consider more 
fully the coupled dynamics of the $A_0$ and fermion sectors, including 
an $x$-dependence for $\Sigma$. Nevertheless, as a first orientation, our 
procedure seems to be reasonably justified.

\section*{Acknowledgments}
C.~D.~F. acknowledges the kind hospitality of the Abdus Salam ICTP (Italy) 
and the High Energy Group of the University of Oxford (U.K.) where part of 
this work  was done. He also acknowledges financial support from CONICET 
and ANPCyT (Argentina). 
This work was also supported by The British Council and Fundaci\'on
Antorchas.

\newpage

\section*{Appendix : Inclusion of the spatial components  of the gauge field.}

In the   SDE's approach, one    usually neglects the  dynamics of  the
spatial components of the gauge  field. This produces  an error in the
result of the self-consistent gap equation.   In the evaluation of the
effective action for $A_0$, on the  other hand, the spatial components
of   the   vacuum   polarization   tensor  yield    a    wave function
renormalization for the spatial components of $A_\mu$.

Let us call $\delta S_{eff}$ the contribution arising from the integration 
over the spatial modes ${\vec a}$ in (\ref{sa}). It is quite straightforward
to see that
$$
e^{-\delta  S_{eff}}\;=\;     \int   {\cal  D}\vec     a   \exp\left\{
-\sum_{n=-\infty}^{+\infty}   \frac{1}{4}  \int  \frac{d^2k}{(2\pi)^2}
{\tilde F}_{jk}(-n,-k) {\tilde F}_{jk}(n,k) \right\}
$$
\begin{equation}
\times  \,  \exp\left\{-\sum_{n=-\infty}^{+\infty}  \frac{1}{2}   \int
\frac{d^2k}{(2\pi)^2} {\tilde a}_i (-n,-k) \Pi_{ij}(n,k)(n,k)  {\tilde
a}_j (n,k) \right\}
\end{equation}
where the tilde denotes Fourier transform. In the static $A_0$ gauge there 
is a  remaining gauge invariance  under space dependent gauge transformations,
which implies that the function $\Pi_{ij}$ is spatially transverse:
\begin{equation}
\Pi_{ij}(n,k) \;=\; \Pi (n,k) (- \Delta \delta_{ij} + \partial_i \partial_j )
\end{equation}

Fixing these gauge field components to be spatially transverse, 
\begin{equation}
e^{-\delta S_{eff}}\;=\; \int {\cal D}\vec a \exp\left\{
-\sum_{n=-\infty}^{+\infty}
\frac{1}{4} \int \frac{d^2k}{(2\pi)^2} {\tilde F}_{jk}(-n,-k)
[1+\Pi (n,k)]
{\tilde F}_{jk}(n,k) \right\} \;.
\end{equation}

Then we may write for the correction to the effective action
\begin{equation}
e^{-\delta  S_{eff}}\,=\, \Pi_{n,{\vec k}} [1+\Pi (n,{\vec k})]^{-1}\, 
\end{equation}
or, equivalently, that there is a correction to the effective potential $V$,
namely, $V \,\to\, V + \delta V$, with
\begin{equation}
\delta V   \,=\, - \sum_n  \int  \frac{d^2k}{(2\pi)^2}\,  \ln  [1 +\Pi
(n,{\vec k}) ]\;.
\end{equation}
In spite  of the fact that this  contribution exists, it is negligible
in the large $N$   limit (note the  absence of  the factor  $N$  which
appears when   integrating  fermions).  Moreover, even    for only one
fermionic flavour,    we checked  that  $\delta  V$   is exponentially
suppressed in comparison with  $V$.
%

\newpage
\input epsf
\epsfbox{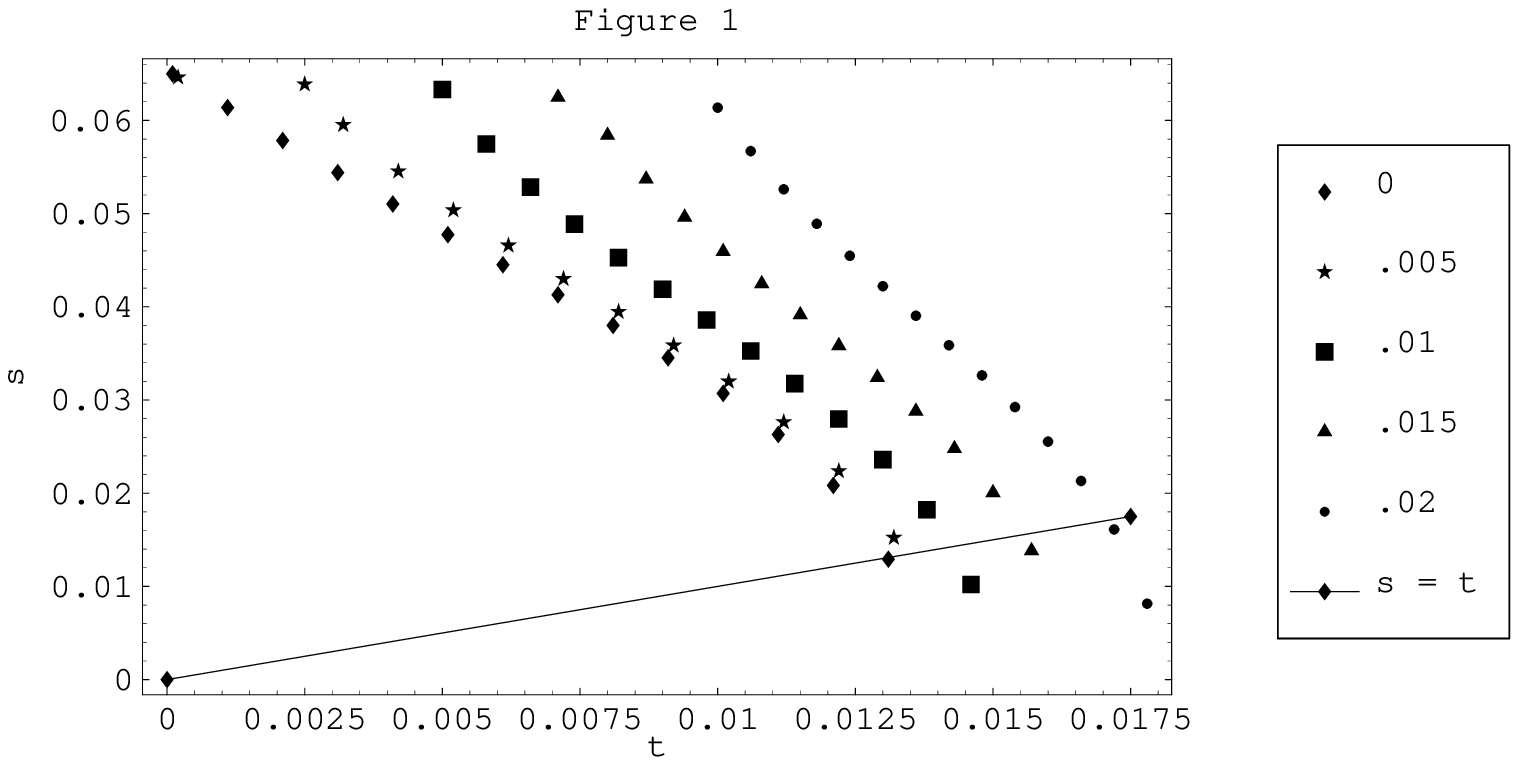}
\epsfbox{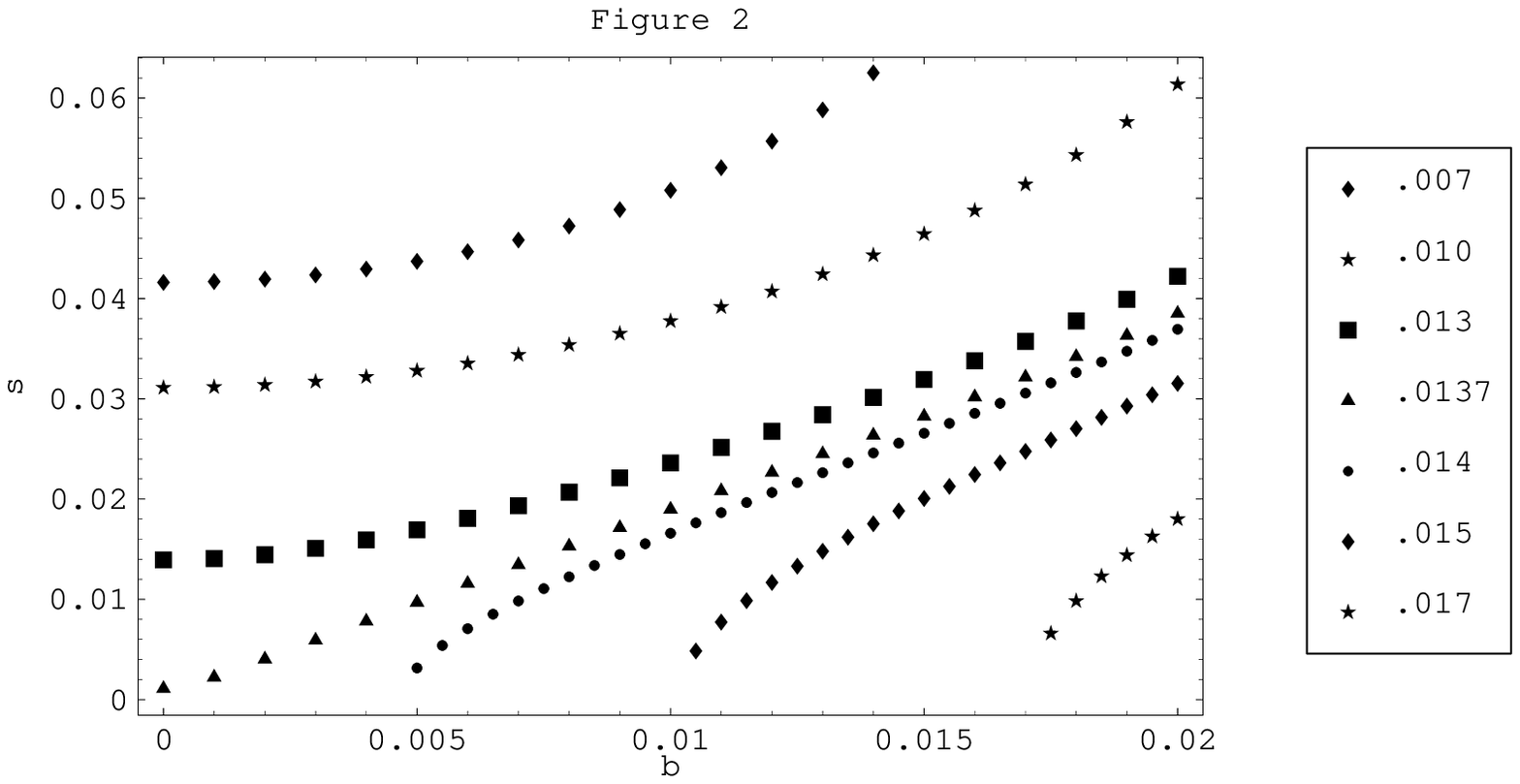}
\newpage
\epsfbox{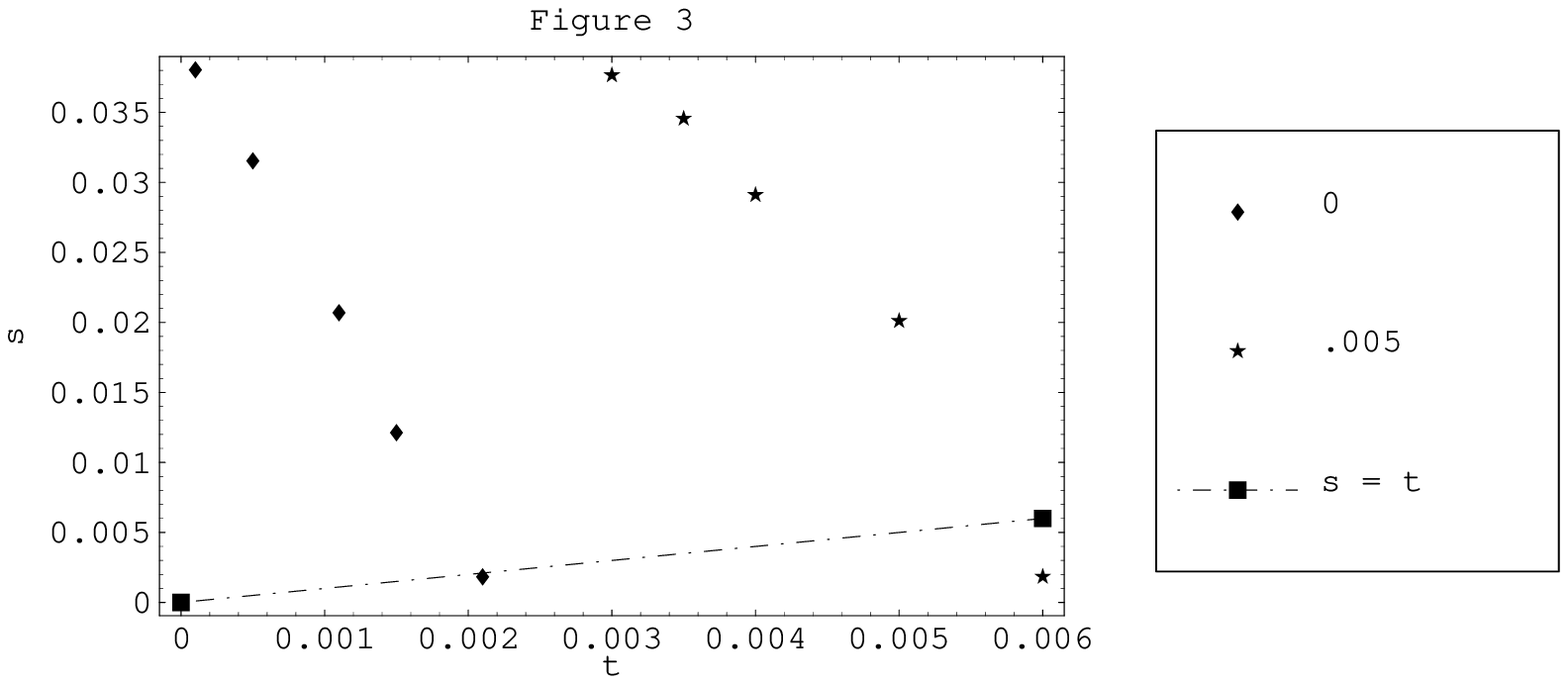}
\newpage
\section*{Figure captions}
\begin{itemize}
\item{\bf Figure 1:} Curves of $s = \frac{\Sigma}{e^2}$ as a function 
of $t = \frac{T}{e^2}$ for various values of 
$b = \frac{{\tilde A}_0}{e}$, for the case $N = 1$. 
\item{\bf Figure 2:} Curves of $s = \frac{\Sigma}{e^2}$ as a function 
of $b = \frac{{\tilde A}_0}{e}$ for various values of 
$t = \frac{T}{e^2}$, for the case $N = 1$. 
\item{\bf Figure 3:} Curves of $s = \frac{\Sigma}{e^2}$ as a function 
of $t = \frac{T}{e^2}$ for two values of 
$b = \frac{{\tilde A}_0}{e}$, for the case $N = 2$. 
\end{itemize}
\end{document}